\documentclass[11pt]{article}
\usepackage{latexsym}
\newcommand{\ft}[2]{{\textstyle\frac{#1}{#2}}}
\begin{document}
\topmargin -25mm
\begin{flushright}
\begin{large}
KUL-TH/2001/1\\
\end{large}
\end{flushright}
\vskip 1.8cm
\centerline{\Large \bf Shadow multiplets and superHiggs mechanism}
\vskip 1.2cm 
\centerline{\bf Davide Fabbri$^a$, Pietro Fr\'e$^b$}
\vskip .8cm \centerline{\sl $^a$ Instituut voor Theoretische Fysica 
- Katholieke Universiteit Leuven}
\vskip .1cm \centerline{\sl Celestijnenlaan 200D B-3001 Leuven, Belgium}
\vskip .1cm \centerline{\sl $^b$ Dipartimento di Fisica Teorica,
Universit\'a di Torino} 
\vskip .1cm \centerline{via P.Giuria 1, I-10125 Torino}
\vskip .1cm \centerline{Istituto Nazionale di Fisica Nucleare (INFN)
- Sezione di Torino, Italy}
\vskip 1cm
\begin{abstract}
We discuss a general feature of Freund Rubin compactifications that was
previously overlooked.
It consist in a curious pairing, which we call a \emph{shadow relation},
of completely different (in terms of spin and mass) fields of the
dimensionally reduced theory. Particularly interesting is the case
where the compactification preserves a certain amount of
supersymmetry, giving rise to a shadowing phenomenon between whole
supermultiplets of fields. In particular, there are strong
suggestions about the consistency of a massive truncation of 11D
supergravity to the massless modes of the graviton supermultiplet
plus the massive modes of its \emph{shadow partner}.
\\
This fact has important consequences in the ${\cal N}=2$ and
${\cal N}=3$ cases, which seem to realize respectively a Higgs or
a superHiggs phenomenon. In other words, we are led to reinterpret
the dimensionally reduced theory as a spontaneously broken phase
of some higher (super)symmetric theory.
\end{abstract}
%
%
\section{Introduction}
The search for consistent compactifications of higher dimensional
supergravity is currently an active field of research \cite{CLP},
mostly due to its connection with the effort to embed brane-world
scenarios in a superstring/M-theory contest \cite{DLS}.
\par
It is quite common to consider several maximally supersymmetric
gauged supergravities as coming from Kaluza Klein sphere
compactifications of higher dimensional supergravities admitting
an $AdS\times$Sphere vacuum solution. The examples include the
$S^4$ and $S^7$ compactifications of 11D supergravity and the
$S^5$ compactification of type IIB supergravity. Generally the
lower dimensional theory is obtained by \emph{truncating} the
spectrum of the higher dimensional supergravity to the massless
sector, namely by setting to zero all the infinite towers of
massive Kaluza Klein modes. By doing so, we are left with an
interacting theory for a finite number of lower dimensional
fields, which can be usually recognized as a gauged supergravity.
\par
It is not a trivial task to show that such a truncation is
consistent, which means that all the solutions of the
dimensionally reduced theory are solutions of the higher
dimensional one too (for instance, it took several years to show
the consistency of the $S^7$ reduction of 11D supergravity down to
${\cal N}=8$ $SO(8)$-gauged supergravity in four dimensions
\cite{dewittnicolai1, dewittnicolai2}). Indeed the full non-linear
equations of motion of the higher dimensional supergravity can be
regarded as the equations of motion for the lower dimensional
fields only at the price of admitting a rather complex non-linear
coupling of these fields with each other. This implies that in
principle infinite different products of massless modes could act
as sources for the massive ones. The consistency of the truncation
requires that this does not happen, namely it requires that the
composite currents coupled to the massive modes always contain at
least one of the massive truncated modes.
\par
As we will show, a very simple argument about the second order
nature of the Kaluza Klein differential field equations leads to
the conclusion that particular truncations of 11D supergravity
(but the argument is sufficiently general to admit generalizations
to other supergravity theories) are consistent. This requires the
concept of \emph{shadow relation} between Kaluza Klein fields that
we are going to explain.
%
%
\section{Shadow multiplets in Kaluza Klein
theory}\label{secshadow}
Following the conventions of \cite{ricpie11} the bosonic action of
11D supergravity reads:
\begin{equation}
\label{boslag11} S = {1\over \kappa_{11}^2} \int {\cal R}\, \det\,
V  - {1\over 16 \kappa_{11}^2} \int F \wedge {}^* F - {1\over 96
\kappa_{11}^2} \int F\wedge F\wedge A~,
\end{equation}
where ${\cal R}$ is the scalar curvature, $V^{M}$ ($M=0,\ldots
10$) are the vielbein, $A$ is a three-form and $F$ its four-form
field-strength. Freund-Rubin (FR) compactifications
\cite{FreundRubin} are solutions of the field equations of 11D
supergravity in which the space-time ${\cal M}_{11}$ has the
factorized form
\begin{equation}
\label{11=4+7} {\cal M}_{11}=AdS_4\times {\cal M}^7~.
\end{equation}
The only non-vanishing components of $F$ are the $4$-dimensional
ones: $F_{abcd}=e\, \varepsilon_{abcd}$, where the parameter $e$
sets the scale for both the 4-dimensional and the $7$-dimensional
cosmological constant (also ${\cal M}^7$ must be an Einstein
space):
\begin{equation}
\label{cosmol4e7} {\cal R}_{ab} = -24\, e^2\, \eta_{ab}~,\hskip
0.8cm {\cal R}_{\alpha\beta} = 12\, e^2\, \eta_{\alpha\beta}~.
\end{equation}
Greek letters $\alpha,\beta,\ldots$ are reserved to flat seven
dimensional indices, while Latin letters $a,b\ldots$ stand for
four dimensional flat indices. We denote by $x$ the coordinates of
four dimensional space, while $y$ are the coordinates on the
compact manifold ${\cal M}^7$.
\par
The fluctuations of the eleven-dimensional fields around such a
background can be expanded into harmonics on the compact
$7$-manifold, namely eigenmodes of proper $7$-dimensional
covariant operators, and the linearized $4$-dimensional field
equations can be suitably diagonalized. The resulting general
formulae were derived in \cite{spectfer,bosmass} and organized
into a systematic way in \cite{univer}. Let us recall the
notations and the main results of that paper. For the fluctuations
$h_{MN}$ of the metric one sets
\begin{eqnarray}
h_{ab}\left(x,y\right) &=& \Big( h_{ab}^I\left(x\right)
    - \frac{3}{M_{(0)^3}+32}D_{(a}D_{b)}
    \left[(2+\sqrt{M_{(0)^3}+36}\,) S^I\left(x\right)\right.
\nonumber\\ &+& \left.
(2-\sqrt{M_{(0)^3}+36}\,)\Sigma^I\left(x\right)\right]
    + \ft54 \delta_{ab}
    \left[(6-\sqrt{M_{(0)^3}+36}\,)S^I\left(x\right)\right.
\nonumber\\ &+& \left.
(6+\sqrt{M_{(0)^3}+36}\,)\Sigma^I\left(x\right)\right]
    \Big) \, Y^I \left(y\right)\,,
\label{kkexpansion}\\ h_{a\beta}\left(x,y\right) &=&
\big[(\sqrt{M_{(1)(0)^2}+16}-4)A_a^I\left(x\right) \nonumber\\ &+&
(\sqrt{M_{(1)(0)^2}+16}+4)W_a^I\left(x\right)\big] \, Y^I_\beta
    \left(y\right) \,,
\label{kkexpansion2}\\ h_{\alpha\beta}\left(x,y\right) &=&
\phi^I\left(x\right) Y^I_{(\alpha\beta)}\left(y\right) -
    \delta_{\alpha\beta}
    \left[(6-\sqrt{M_{(0)^3}+36}) S^I\left(x\right)\right.
\nonumber \\ &+& \left.  (6+\sqrt{M_{(0)^3}+36})\Sigma^I
    \left(x\right)\right] \, Y^I\left(y\right)~.
\label{kkexpansion3}
\end{eqnarray}
For the fluctuations $a_{MNR}$ of the three form field, one has
\begin{eqnarray}
\label{kkexpansionA} a_{abc}\left(x,y\right) &=& 2 \,
\varepsilon_{abcd} \,D_d (S^I\left(x\right)+
    \Sigma^I\left(x\right)) Y^I\left(y\right) \,,
\nonumber \\ a_{ab\gamma}\left(x,y\right)
             &=& \ft23 \, \varepsilon_{abcd} \,
            (D_c A_d^I\left(x\right) + D_c W_d^I\left(x\right))\,
        Y_\gamma^I\left(y\right) \,,
\nonumber \\ a_{a\beta\gamma}\left(x,y\right)
             &=& Z_a^I\left(x\right) Y^I_{[\beta\gamma]}\left(y\right) \,,
\\
a_{\alpha\beta\gamma}\left(x,y\right)
            &=& \pi^I\left(x\right) Y^I_{[\alpha\beta\gamma]}\left(y\right)~.
\end{eqnarray}
Finally, for the fluctuations of the gravitino field,
\begin{eqnarray}
\psi_a\left(x,y\right)
            &=& \Big(
                \chi_a^I\left(x\right) +
        \frac{\ft47 M_{(1/2)^3}+8}{M_{(1/2)^3}+8}
                \big[D_a \lambda_L^I\left(x\right) \big]_{3/2}
\nonumber\\ &-&    (6+\ft37 M_{(1/2)^3})\gamma_5\gamma_a
\lambda_L^I \left(x\right)
           \Big) \, \Xi^I\left(y\right) \,,
\label{kkexpansionpsi}\\ \psi_\alpha(x,y) &=&
\lambda_T^I\left(x\right) \Xi_\alpha^I\left(y\right) +
                \lambda_L^I\left(x\right)
                \big[ \nabla_\alpha \Xi ^I\left(y\right) \big]_{3/2}~.
\label{kkexpansionlam}
\end{eqnarray}
The harmonics on ${\cal M}^7$ are  grouped into seven infinite
towers corresponding to as many irreducible representations of the
tangent group $SO(7)$ that appear in the decomposition $4\oplus 7$
of eleven dimensional tensors and spinors. Since $SO(7)$ has rank
3, its {\sl irreps} are labeled by three numbers $[\lambda _1
,\lambda _2 ,\lambda _3]$ that we take to be the Young
labels\footnote{ That is, for bosonic tensors with symmetry
represented by a Young tableaux, $\lambda_i$ is the number of
boxes in the $i$-th row of the tableaux. For gamma-traceless
irreducible spinor tensors $\lambda^i$ is $1/2$ plus the number of
boxes.}.
\par
Measuring everything in units of the Freund-Rubin scale,
\emph{i.e.} setting $e=1$, the masses of the $4$-fields appearing
in the Kaluza Klein expansion (eq.s
\ref{kkexpansion}-\ref{kkexpansionlam}) are expressed in terms of
the eigenvalues $M_{[\lambda _1 ,\lambda _2 ,\lambda _3]}$ of the
appropriate $7$-operators as follows:
\begin{eqnarray}
m_h^2 & = & M_{\left(0\right)^3}\,, \label{gmassh} \\
m_{\Sigma}^2&=& M_{\left(0\right)^3} +176+24\sqrt{
M_{\left(0\right)^3}+36}\,, \label{gmassigma}\\ m_S^2 &=&
M_{\left(0\right)^3} +176-24\sqrt{
M_{\left(0\right)^3}+36}\,,\label{gmassesse}\\ m_{\phi}^2 &=&
M_{\left(2\right)\left(0\right)^2} \,,\label{gmassfi}\\ m_{\pi}^2
&=& 16\left( M_{\left(1\right)^3}-2\right)\left(
M_{\left(1\right)^3}-1\right) \,,\label{gmasspi}\\ m_W^2 &=&
M_{\left(1\right)\left(0\right)^2} + 48 + 12
\sqrt{M_{\left(1\right)\left(0\right)^2}+16}\,,\label{gmassW}\\
m_A^2 &=& M_{\left(1\right)\left(0\right)^2} + 48 - 12
\sqrt{M_{\left(1\right)\left(0\right)^2}+16}\,,\label{gmassA}\\
m_Z^2 & = & M_{\left(1\right)^2\left(0\right)} \,,\label{gmassZ}\\
m_{\lambda_L} & = & -\left( M_{\left(1\over 2\right)^3}
+16\right)\,, \label{gmasslamL}\\ m_{\lambda_T} & = &
M_{\left(3\over 2\right)\left(1\over
2\right)^2}+8\,,\label{gmasslamT}\\ m_{\chi} & = & M_{\left(1\over
2\right)^3}\,.\label{gmasschi}
\end{eqnarray}
The AdS relation between the mass $m_{(s)}$ and the rest energy
$E_{(s)}$ of a spin $s$ particle which, in the alternative
three-dimensional conformal interpretation of the $SO(2,3)$ group,
translates into a relation with the scale dimension $\Delta=E$ of
the corresponding primary conformal field (see \cite{Osp} for
further details), is given by\footnote{The reader should be
careful in comparisons with other papers and take into account
that the definition of mass utilized in this paper is that of
supergravity \cite{castdauriafre}. Specifically the mass squared
of scalars is defined as the deviation from a conformal invariant
equation, the mass of the gravitino is defined as the deviation
from a Rarita Schwinger equation with supersymmetry, the mass
squared of a spin one field is defined as the deviation from a
gauge invariant equation.}:
\begin{eqnarray}
m_{\left(0\right)}^2&=&16\left(E_{\left(0\right)}-2\right)
\left(E_{\left(0\right)}-1\right) \,,\nonumber\\ |m_{\left(1\over
2\right)}|&=&4E_{\left(1\over 2\right)}-6\,,\nonumber\\
m_{\left(1\right)}^2&=&16\left(E_{\left(1\right)}-2\right)
\left(E_{\left(1\right)}-1\right) \,,\nonumber\\ |m_{\left(3\over
2\right)}+4|&=&4E_{\left(3\over 2\right)}-6~. \label{massenergy}
\end{eqnarray}
\paragraph{The shadowing phenomenon}
Having established the conventions on the Kaluza Klein expansion
of the eleven dimensional fields, we are now able to discuss the
shadowing phenomenon in details. As it can be seen from equations
(\ref{kkexpansion}-\ref{kkexpansionlam}), there are infinitely
many couples of fields which share the same harmonics. The masses
of the fields in each couple are related through equations
(\ref{gmassh} -\ref{gmasschi}) to the same eigenvalue of the
corresponding harmonic. This means that fields of different type,
spin and mass are nevertheless linked by a relation which
determines the mass of the one as a function of the other, as
shown in table \ref{shadowcouples}. Furthermore, each field
belongs to the same irrep of the isometry group of ${\cal M}^7$ as
the shadow partner, which is just the conjugate representation of
the associated harmonic.
\renewcommand{\arraystretch}{1}
\begin{table}
\begin{center}
{\small
\begin{tabular}{cc}
\hline\hline {\rm shadow couples} & {\rm mass relation}\\ \hline
\\
$(h^I_{ab},\ S^I)$ & $m^2_S=m^2_h+176-24\sqrt{m^2_h+36}$\\
\\
$(h^I_{ab},\ \Sigma^I)$ &
$m^2_\Sigma=m^2_h+176+24\sqrt{m^2_h+36}$\\
\\
$(A^I_a,\ W^I_a)$ & $m^2_w=m^2_A+144-24\sqrt{m^2_A+4}$\\
\\
$(\psi^I_a,\ \lambda^I_L)$ & $m_{\lambda_L}=-m_\psi-16$\\
\\
\hline\hline
\end{tabular}
}
\end{center}
\caption{Shadow couples. The mass of each field is related to that
of the shadow partner.} \label{shadowcouples}
\end{table}
\paragraph{Examples}
We can illustrate the basic structure of this mechanism with some
simple examples that will be quite relevant in our subsequent
discussion.
\par
Let us observe that the same scalar harmonic $Y^I(y)$ is
associated both to the graviton field $h^I_{ab}(x)$ of eq.
(\ref{kkexpansion}) and to the scalar field $\Sigma^I(x)$ of eq.
(\ref{kkexpansion2}). Using the mass relations (\ref{gmassh}
-\ref{gmasschi}), we see that the shadow scalar of a ``parent''
graviton has mass
\begin{equation}
  m_\Sigma^2 =m_h^2 +176 +24 \sqrt{m_h^2+36}
\label{shasca}
\end{equation}
In the case of the massless graviton, $m_h^2=0$, corresponding to
the constant harmonic $Y=1$, its shadow scalar $\Sigma$ has
(squared) mass
\begin{equation}
  m^2_\Sigma = 320~.
\label{massigma}
\end{equation}
Using eq.s (\ref{massenergy}) this implies
\begin{equation}
  E_\Sigma = 6~.
\label{Esigma}
\end{equation}
We conclude that in every $AdS_4 \times {\cal M}^7$
compactification there is always a \emph{universal} scalar mode of
conformal dimension $E=6$ that is the shadow of the graviton. Its
geometric origin is apparent from the second of equations
(\ref{kkexpansion}): it is just the \emph{breathing mode}
corresponding to an overall dilatation of the internal manifold
${\cal M}^7$.
\par
As it follows by inspection of eq.s (\ref{kkexpansion2}), to the
same vector harmonic $Y^I_\alpha $ we associate two vector fields:
one, $A_a$, with mass given by eq. (\ref{gmassA}), the other,
$W_a$, with mass given by eq. (\ref{gmassW}). This is due to the
fact that the linearized field equation for a $4$-vector
associated to an harmonic of given $M_{(1)(0)^2}$ is a second
order differential equation. Hence we have two independent
solutions rather than one. The very initial idea of Kaluza Klein
theory is that the isometries of the internal manifold give rise
to massless gauge fields in the compactified $4$-dimensional
theory. When we start from 11D M-theory there is a further aspect.
Indeed to each Killing vector of the internal compact manifold we
associate two rather than one vector fields. In addition to the KK
massless gauge boson, we have its shadow massive vector. It also
belongs to the adjoint representation of the isometry group, and
it has fixed mass and dimension:
\begin{equation}
  m^2_W = 192 \, \Rightarrow \, E_W =5~.
\label{emme2w}
\end{equation}
\par
Finally, let us consider one more example involving fermionic
fields. Comparing eq.s (\ref{kkexpansionpsi}) and
(\ref{kkexpansionlam}), we see that the same spinor harmonic
$\Xi^I$ appearing in the expansion of the gravitino $\psi_a(x,y)$
has as coefficients of the expansion both a spin-3/2 $\psi^I_a(x)$
and a longitudinal spin-1/2 field $\lambda_L(x)^I$. Using eq.s
(\ref{gmasslamL},\ref{gmasschi}) the relation between the masses
of the spin-$3/2$  and spin-$1/2$ modes pertaining to the same
harmonic is:
\begin{equation}
  m_\psi=-m_{\lambda_L} -16~.
\label{mchimlam}
\end{equation}
Applying eq. (\ref{mchimlam}) to the case $m_{\lambda_L}=0$, we
see that each massless spin $\ft 12$ particle of this type
generates a shadow massive gravitino with mass:
\begin{equation}
  m_\psi = -16 \, \Rightarrow \, E_\psi= \frac 9 2 ~.
\label{mchi92}
\end{equation}
Conversely, every massless gravitino produces a shadow massive
spin-1/2 field with mass:
\begin{equation}
  m_{\lambda _L} = -16 \, \Rightarrow \, E_{\lambda _L}= \frac {11}2~.
\label{mlam112}
\end{equation}
\paragraph{Supersymmetry and shadow multiplets}
Let us consider unitary irreducible representations of the
superalgebra $Osp({\cal N}\vert 4)$, {\it i.e.} the supersymmetric
extension of $SO(2,3)$ with ${\cal N}$ supercharges. Each of them
is a  supermultiplet, represented by a suitable constrained
superfield, that contains fields whose spins $s$ and dimensions
$E$ are related to each other.
\par
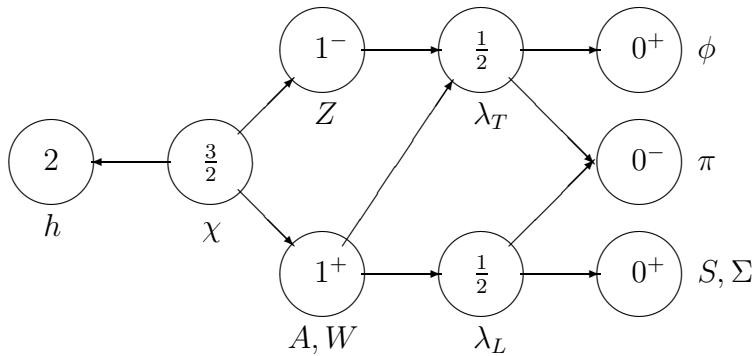
\begin{figure}
\centering
\begin{picture}(268,200)
\put (15,100){\circle{30}} \put (12,97){\shortstack{\large{$2$}}}
\put (12,72){\shortstack{\large{$h$}}} \put (60,100){\vector
(-1,0){30}} \put (75,100){\circle{30}} \put
(72,97){\shortstack{\large{$3\over 2$}}} \put
(72,72){\shortstack{\large{$\chi$}}} \put (85.61,110.61){\vector
(1,1){21.22}} \put (85.61,89.39){\vector (1,-1){21.22}} \put
(117.43,142.43){\circle{30}} \put
(114.43,139.43){\shortstack{\large{$1^-$}}} \put
(114.43,114.43){\shortstack{\large{$Z$}}} \put
(117.43,57.57){\circle{30}} \put
(114.43,54.57){\shortstack{\large{$1^+$}}} \put
(104.43,29.57){\shortstack{\large{$A,W$}}} \put
(132.43,142.43){\vector (1,0){30}} \put (132.43,57.57){\vector
(1,0){30}} \put (125.04,68.18){\vector (2,3){42}} \put
(177.43,142.43){\circle{30}} \put
(174.43,139.43){\shortstack{\large{$1\over 2$}}} \put
(174.43,114.43){\shortstack{\large{$\lambda_T$}}} \put
(177.43,57.57){\circle{30}} \put
(174.43,54.57){\shortstack{\large{$1\over 2$}}} \put
(174.43,29.57){\shortstack{\large{$\lambda_L$}}} \put
(192.43,142.43){\vector (1,0){30}} \put (188.04,131.82){\vector
(1,-1){32.5}} \put (188.04,68.18){\vector (1,1){32.5}} \put
(192.43,57.57){\vector (1,0){30}} \put
(237.43,142.43){\circle{30}} \put
(234.43,139.43){\shortstack{\large{$0^+$}}} \put
(259.43,139.43){\shortstack{\large{$\phi$}}} \put
(237.43,100){\circle{30}} \put
(234.43,97){\shortstack{\large{$0^-$}}} \put
(259.43,97){\shortstack{\large{$\pi$}}} \put
(237.43,57.57){\circle{30}} \put
(234.43,54.57){\shortstack{\large{$0^+$}}} \put
(259.43,54.57){\shortstack{\large{$S,\Sigma$}}}
\end{picture}
\caption{\emph{``Internal''} supersymmetry relations between the
harmonics of Kaluza Klein fields: for each pair of fields linked
by an arrow the corresponding towers of harmonics can be related
to each other by multiplications with a Killing spinor $\eta^A$.}
\label{molecola}
\end{figure}

These relations hint to a sort of mirror image of the
supersymmetry algebra that is realized on the internal compact
manifold ${\cal M}^7$. This idea was thoroughly analyzed in
\cite{univer} and traced back to the existence of (commuting)
Killing spinors $\eta^A$ ($A=1,\ldots {\cal N}$), where ${\cal N}$
is the number of preserved supersymmetries of the $AdS_4 \times
{\cal M}^7$ compactification one considers. By means of the
Killing spinors $\eta^A$, to each harmonic $Y$ that is an
eigenmode of a bosonic $7$-operator one can associate another
fermionic harmonic $\Xi$ that is an eigenmode of a fermionic
$7$-operator with suitably related eigenvalues. These pairs of
related harmonics were explicitly constructed in \cite{univer} and
follow the schematic pattern given in fig. \ref{molecola}.
\par
As already stressed fifteen  years ago in \cite{univer}, these
relations are differential geometric identities on the compact
$7$-manifold ${\cal M}^7$ that are required by consistency with
the structure of UIR.s of the superconformal group $Osp({\cal
N}\vert 4)$. Because of this, it may at first sight appear that
the universal mass relations analyzed in \cite{univer} do not
contain further physical information besides the implications of
supersymmetry. However, this is not the case, because of another
aspect of eq.s (\ref{gmassh}-\ref{massenergy}), whose consequence
is precisely the existence of shadow multiplets.
\par
The point is that the relations (\ref{gmassh}-\ref{gmasschi})
between masses (or conformal weights) and eigenvalues of the
internal Laplacian is quadratic rather than linear. Indeed the
same harmonic always  plays a double role since it appears in the
expansion of two quite different Kaluza Klein fields. Combined
with the supersymmetry relations produced by Killing spinors, this
has the curious consequence that each supersymmetry multiplet of
the Kaluza Klein spectrum is associated with another one made, so
to say, by the second roots of the quadratic relations.
%
%
\section{Consistency of the shadow truncation}
Let us consider the $7$-dimensional equation satisfied by the
scalar harmonics associated to the spin two Kaluza Klein fields
$h^I_{ab}$:
\begin{equation}
  \Box_7 Y = D^\mu D_\mu Y = M_{(0)^3} Y\,.
\end{equation}
The compactness of ${\cal M}^7$ implies the positivity of $\Box_7$
and, in particular, the fact that the lightest spin two field,
i.e. the massless graviton $h^0_{ab}$ is associated to the
constant harmonic $Y=1$. Together with the killing spinor
relations of table \ref{molecola}, this implies that all the
harmonics associated to the fields of the massless graviton
supermultiplet are given by some product of killing spinors and
constant covariant tensors. The same is true, by definition, for
the harmonics associated to the fields of the shadow multiplet
(i.e. the shadow partner of the graviton multiplet). This implies
that more general harmonics, associated to the other massive
fields, cannot be obtained as products of the only Killing spinor
harmonics. Hence in the complete (non-linear) $4$-dimensional
equations of motion, there are no non-vanishing currents (i.e.
composites of the only massless and shadow fields) that could act
as sources for the massive truncated fields. The conclusion is
that the truncation of the Kaluza Klein modes to the massless
graviton multiplet plus the shadow sector has to be consistent,
independently from the specific compact manifold ${\cal M}^7$.
\par
Actually it is worth to stress that this is not a rigorous proof
of consistency because, in principle, we do not have a definitive
argument to exclude the existence of at least one field, among the
higher massive modes, whose harmonic is made by a product of
constant tensors and Killing spinors. If this were the case, the
consistent truncation could require not to switch off this field.
But this possibility seems very unrealistic. First of all because,
as we have seen, the harmonics of the fields in the same
supermultiplet would be related to this one by products with
Killing spinors. So they all would be of the same kind and we
should deal with a whole new massive multiplet to retain in the
consistent truncation. Second, because the shadow multiplet fields
have a really universal geometrical meaning, while the other
massive ones are related to the specific features of the
compactification manifold. It seems therefore really plausible
that there is a deep link between the most general kind of
multiplet of a supergravity theory (the massless graviton) and
another universal multiplet (its shadow).
%
%
\section{The superHiggs phenomenon}
The conclusion about the consistency of the truncation of any
supersymmetric compactification of 11D supergravity to the
massless plus the shadow sectors implies that the dimensionally
reduced theory is a consistent coupling of some gauged
supergravity to a proper multiplet of massive fields. In
particular, in the case of ${\cal N}=3$ supersymmetric
compactifications, independently on the internal manifold ${\cal
M}^7$, there has to be a consistent coupling of a whole massive
gravitino multiplet to the ${\cal N}=3$ $SO(3)$-gauged
supergravity in four dimensions. Indeed the shadow partner of the
${\cal N}=3$ graviton multiplet, displayed in table
\ref{ushorgrav}, is a massive gravitino multiplet.
\begin{table}
  \begin{center}
{\small
 \begin{tabular}{ccc|cc}
      \hline\hline
      Spin & Energy & $SO(3)$-Isospin  & Field & Harmonic\\
      \hline &&&&\\
      $2$ & $3$ & $0$ & $h_{ab}$ & $Y=1$\\
      &&&&\\
      $\ft32$ & $\ft52$ & $1$& $\chi^A_a$ &
           $\Xi^A =\eta^A$ (Killing spinor) \\
      &&&&\\
      $1$  & $2$ & $1$ &$A_a^A$&
      $Y^A_\alpha =\ft 12 \, \epsilon^{ABC} \,
      {\bar \eta}^B \, \gamma_\alpha  \eta^C \equiv k^A_\alpha $
      \\
      &&&&\\
      $\ft12$  & $\ft32$ & $0$ & $ \lambda_L$ & $\Xi=\ft 13
      \,\epsilon^{ABC}\, \gamma_\alpha \, \eta^A \,{\bar \eta}^B \,
      \gamma^\alpha  \eta^C $\\
      &&&&\\
      \hline\hline
    \end{tabular}
    }
    \caption{The massless ${\cal N}=3$ graviton multiplet and
    its Kaluza Klein origin.}
    \label{ushorgrav}
  \end{center}
\end{table}
The only way to introduce such a multiplet in a consistent way is
through the super-Higgs phenomenon\footnote{analogously, in the
${\cal N}=2$ case, where the shadow partner of the massless
graviton is a massive vector multiplet, we deal with a simple
Higgs phenomenon}. Hence we are led to the conclusion that any
${\cal N}=3$ dimensional reduction of 11D supergravity on a
background of the form $AdS_4\times{\cal M}^7$ actually is a
broken phase of some gauged ${\cal N}=4$ supergravity.
\par
The ${\cal N}=4\to 3$ partial supersymmetry breaking in
four-dimensional supergravity is well-known in the literature
\cite{wageman}. It can be realized by coupling three vector
multiplets to the $SO(4)$-gauged ${\cal N}=4$ supergravity and
giving a proper {\it vev} to four scalars which preserve the
$SO(3)$ $R$-symmetry of the ${\cal N}=3$ vacuum. As explicitly
shown in \cite{noi}, there is an upper bound on the mass acquired
by the broken gravitino, which is a function of these {\it vev}s
(the moduli of the partial breaking). The rest energy of this
gravitino is given by:
\begin{equation}
  E_{\psi}=\frac{3}{2}+\sqrt{\frac{8+\rho^4
  -\rho^2 (8+\rho^2)\cos 4 \theta}
  {8+\rho^4-8\rho^2-\rho^4\cos 4 \theta}}\,,
\end{equation}
where the complex variable $\rho e^{i\theta}$, $\rho<1$, partially
parametrizes the partial breaking moduli space. Hence standard
${\cal N}=4$ supergravity poses the upper bound $E_{\psi} < 9/2$.
\par
Now, in the case where the supergravity we are considering is
realized through a dimensional reduction from 11D, the broken
gravitino is nothing but the higher spin component of the shadow
supermultiplet of the compactification. As we have shown in the
third example of section \ref{secshadow} and as it is discussed in
\cite{N010} for a particular compactification, the energy of such
field is just $E_{\psi} = 9/2$, independently from the specific
${\cal M}^7$. This means that there must exist some new way to
realize the ${\cal N}=4\to 3$ supersymmetry breaking, admitting a
broken gravitino energy $E_{\psi}\geq 9/2$, so far not considered
in the literature.
%
%


\begin{thebibliography}{99}

\bibitem{CLP}
H.~Nastase, D.~Vaman and P.~van~Nieuvenhuizen {\it Consistency of
the $AdS_7\times S_4$ reduction and the origin of self-duality in
odd dimensions}, Nucl.Phys. {\bf B581} (2000) 179, hep-th/9911238;
M.~Cvetic, H.~Lu and C.N.~Pope {\it Consistent Kaluza-Klein sphere
reductions}, Phys.Rev. {\bf D62} (2000) 046005, hep-th/0003286;
M.~Cvetic, H.~Lu, C.N.~Pope, A.~Sadrzadeh and T.A.~Tran {\it
Consistent $SO(6)$ reduction of type IIB supergravity on $S^5$},
Nucl.Phys. {B586} (2000) 275, hep-th/0003103.

\bibitem{DLS}
M.J.~Duff, J.T.~Liu and K.S.~Stelle {\it A supersymmetric Type IIB
Randall-Sundrum realization}, hep-th/0007120; M.~Cvetic, H.~Lu and
C.N.~Pope {\it Domain walls and massive gauged supergravity
potentials}, hep-th/0001002.

\bibitem{dewittnicolai1}
B.~de~Wit and H.~Nicolai {\it The embedding of N=8 supergravity
into $D=11$ supergravity}, Nucl.Phys. {\bf B255}, (1985) 29.

\bibitem{dewittnicolai2}
B.~de~Wit and H.~Nicolai {\it The consistency of the $S^7$
truncation in $D=11$ supergravity}, Nucl.Phys. {\bf B281}, (1987)
211.

\bibitem{ricpie11}
R.~D'Auria and P.~Fr\'e {\it Geometric d=11 supergravity and its
hidden supergroup}, Nucl.Phys. {\bf B201} (1982) 101.

\bibitem{FreundRubin}
P.G.O.~Freund and M.A.~Rubin {\it Dynamics of dimensional
reduction}, Phys.Lett. {B97} (1980) 233.

\bibitem{spectfer}
R.~D'Auria and P.~Fr\'e, {\it On the fermion mass-spectrum of
Kaluza Klein supergravity}, Ann.Phys. {\bf 157} (1984) 1.

\bibitem{bosmass}
L.~Castellani, R.~D'Auria, P.~Fr\'e, K.~Pilch,
P.~van~Nieuwenhuizen {\it The bosonic mass formula for
Freund-Rubin solutions of $D$=11 supergravity on general coset
manifolds}, Class.Quant.Grav. {\bf 1} (1984) 339.

\bibitem{univer}
R.~D'Auria and P.~Fr\'e {\it Universal Bose-Fermi mass-relations
in Kaluza Klein supergravity and harmonic analysis on coset
manifolds with Killing spinors}, Ann.Phys. {\bf 162} (1985) 372.

\bibitem{heidenreich}
W.~Heidenreich {\it All Linear Unitary Irreducible Representations
of de~Sitter Supersymmetry with Positive Energy}, Phys.Lett. {\bf
B110} (1982) 461.

\bibitem{frenico}
D.~Freedman and H.~Nicolai {\it Multiplet shortening in
$Osp(N,4)$}, Nucl.Phys. {\bf B237} (1984) 342-366.

\bibitem{multanna}
A.~Ceresole, P.~Fr\'e, H.~Nicolai {\it Multiplet structure and
spectra of $N=2$ supersymmetric compactifications},
Class.Quant.Grav. {\bf 2} (1985) 133-145.

\bibitem{castdauriafre}
L.~Castellani, R.~D' Auria and P.~Fr\`e, {\it Supergravity and
superstrings: A geometric perspective}, World Scientific,
Singapore (1991).

\bibitem{Osp}
D.~Fabbri, P.~Fr\`e, L.~Gualtieri, P.~Termonia {\it $Osp({\cal
N}|4)$ supermultiplets as conformal superfields on $\partial
AdS_4$ and the generic form of ${\cal N}=2$, $d=3$ gauge
theories}, Class.Quant.Grav. {\bf 17} (2000) 55, hep-th/9905134.

\bibitem{wageman}
M.~de~Roo and P.~Wagemans {\it Partial supersymmetry breaking in
N=4 supergravity}, Phys.Lett. {\bf 177} (1986) 352.

\bibitem{noi}
M.~Bill\`o, D.~Fabbri, P.~Fr\`e, P.~Merlatti and A.~Zaffaroni {\it
Shadow multiplets in $AdS_4/CFT_3$ and the super-Higgs mechanism:
hints of new shadow supergravities}, Nucl.Phys. {\bf B591} (2000)
139, hep-th/0005220.

\bibitem{N010}
M. Bill\`o, D. Fabbri, P. Fr\`e, P. Merlatti, A.~Zaffaroni {\it
Rings of short ${\cal N}=3$ superfields in three dimensions and
M-theory on $AdS_4\times N^{010}$}, hep-th/0005219.

\end{thebibliography}
\end{document}